\documentclass{article}

\usepackage{graphicx}
\usepackage{hyperref}

\begin{document}



\title{The AI Accountability Ecosystem in the Era of Large Language Models}

\author{Chris Percy\thanks{University of Warwick, Coventry, CV4 7AL, UK} \and Artur d'Avila Garcez\thanks{City St George's, University of London, London, EC1V 0HB, UK}}


\maketitle

\begin{abstract} 
This article reviews and updates a conceptual framework for accountability in AI based on \textit{accountability ecosystems}. We update the framework in light of relevant developments since the release of Large Language Models for general public use in 2023. We propose three interlinked updates to the original AI accountability ecosystem: (i) reorienting the accountability ecosystem to AI infrastructure and supply chains, (ii) providing greater emphasis on outcomes monitoring and identification of issues that support decentralized system improvement, and (iii) incorporating end-user accountability given the new risks of unpredictability of language models \textit{in-the-wild}. Collectively, these updates mark a shift towards accountability as distributed, continuous, and institutionalized, away from a system in which frontier AI applications can be modeled as discrete products controlled by single identifiable actors with industry-specific oversight.
\end{abstract}


\sloppy

\section{Introduction}

This article reviews and updates a conceptual framework for accountability in AI in light of recent developments in Large Language Models (LLMs) with agentic AI. The framework introducing an \textit{accountability ecosystem in AI} predated the Generative AI revolution with LLMs.$^{1}$ Nevertheless, it contained relevant predictions about today's AI landscape. Written in 2021, it concluded that ``if we collectively fail to make progress towards these [accountability] goals, towards measurable explainability and implementing these accountability mechanisms, we can look forward to immature regulation operating in an unbalanced ecosystem, being either ineffective or overwhelming, and continued frustration from a civil society that lacks constructive channels through which to direct their anger and to help shape AI practice''.$^{1}$ 

When the \textit{accountability ecosystem in AI} idea was introduced five years ago, the infrastructure layer for AI was more benign, with risks concentrated more in application development than underlying infrastructure. Those building AI applications did draw on an infrastructure layer, but often limited to cloud services for processing their own data, standard libraries to implement well-understood algorithms, or broad-based online communication infrastructures. In the early 2020s, it was reasonable for an accountability ecosystem to be built around the technical team or company developing the application, even if multiple stakeholder layers and mechanisms were required to hold the developers adequately to account.

Today's world is very different. The era of Large Language Models created a centralized infrastructure layer driving AI applications that is far from benign. Frontier models are revised without forward-looking schedules, producing new capabilities and limitations that are only weakly understood by their owners, let alone their API-users. Design choices are more art than science, involving opaque decisions on training set corpora, algorithm optimization, and applicable AI principles. Four years ago, Reinforcement Learning with Human Feedback (RLHF) was promoted as the main technological innovation behind LLMs. Now, the technical limitations and ethical woes of RLHF have become obvious given the failures to obtain reliability and accountability.

With the above developments, the weaknesses in our collective AI accountability ecosystem remain all too apparent. This article explores what a strong ecosystem might look like in today's context. We argue for three interlinked adjustments to the original ecosystem proposal. First, reorient the accountability ecosystem around supply chains, rather than around the development team building a given application, so as to recognize the importance of assurance markets and benchmark validation across a delivery stack. Second, apply greater emphasis on outcomes monitoring and identification of issues by stakeholders in ways that support decentralized system improvement. Third, explicitly incorporate user accountability and safeguards, given the new risks raised by recursive self-improvement and system unpredictability with \textit{sandbox escapes}. The remaining elements of the original accountability ecosystem framework continue to be valid, with an emphasis on a tiered approach to risk management, taking into account the contributions and interactions between corporate actors, their market counterparts, civil society, and government. At the same time, stronger tools are necessary to deliver on the requirements of the original proposal, as we will explore in a brief discussion on neurosymbolic and explainable AI later in the paper.

Collectively, the above three adjustments mark a shift towards accountability as a distributed, continually monitored and institutionalized process, away from a situation where frontier AI applications can be treated as discrete products controlled by single identifiable actors with industry-specific oversight. Where the most significant risks occur in infrastructure layers, the accountability approach might draw inspiration from such areas as communication network infrastructure or finance. Supplier licensing, model design certification, independent safety bodies, and incident reporting systems have a larger role to play than internal and external auditing directed at corporate API-users. 

At the same time, the ecosystem perspective from 2021 remains key to the collective approach to AI accountability: no single mechanism is sufficient. Multiple societal layers all have a role to play, from technical teams and companies through to markets, civil societies, and governments. Explainability, transparency, and informed participation also remain essential principles underpinning accountability, now magnified by problems of large-scale data security and privacy. The focus on operationalization from 2021 becomes even more important as AI technologies are already too diffusely embedded for adherence to principles to be trusted or directly monitored.$^{2}$ Converting principles into practice continues to be important, as was proposed in 2021 using \textit{nested operational procedures}, from local decision process documentation and internal audits to external audits and accreditation.$^{1}$  

This article does not question the current and potential contributions of LLMs, nor do we argue for turning back the clock. We acknowledge the value of LLMs as an important engineering achievement with great potential to derive productivity gains across the global economy. The question is how to unlock these contributions in a safe and sustainable way. In what follows, we take the three proposed adjustments in turn, beginning by setting out a selection of the distinctive issues introduced by the LLM era and summarizing the AI ecosystem approach upon which this work builds. This paper provides a conceptual discussion of ecosystem issues, recognizing that significant further work is needed to operationalize it adequately.

\section{New Accountability Issues in the era of LLMs}
In an unregulated area of knowledge such as computer systems, or in a marketplace without accountability, a system like ChatGPT can be made available to billions of users for free without stringent quality checks, ethical certification, or reliability guarantees. In some corners, this is seen as creating a healthy disruption in the market. Google certainly felt that disruption when OpenAI released ChatGPT in late 2022. The ease with which internet searches were enabled via system interactions in natural language put Google's keyword-driven internet advertising business model in jeopardy. What followed was a Big Tech race to the bottom with companies competing fiercely for user engagement, lawsuits, a lack of transparency, and a normalization of the idea that AI will \textit{make stuff up}, in some cases with serious consequences.$^{3}$  

\subsection{\textit{LLM hallucinations have not gone away}} 
Imperfect reliability is a fact of life outside the self-contained world of axiomatic proofs and high-assurance industries. Complex socio-technical systems are very hard to verify formally and humans make mistakes. However, incentives, oversight structures, and the use of formal methods have been shown to promote error correction processes that can be deployed to improve reliability with costs proportionate to a given use case.$^{4}$ In cases with high-impact errors, such as air travel or infrastructure security, multiple layers of automated and human oversight reduce mistakes to acceptable levels, where what counts as acceptable is continually negotiated between regulators, stakeholders, and users.

Today's LLM architectures have not yet mastered this proportionality. LLM ``hallucinations'' have reduced in recent years, but it is not yet possible to invest in general-purpose AI infrastructure where errors are reduced to a pre-determined level. The only current way to achieve this is via significant domain-specific harnesses or human-in-the-loop checks, which largely undo the productivity gains of AI and place unrealistic pressure on human monitors. Manually finding errors in a stream of AI output is tantamount to looking for discolored pixels in a photo, especially with AI algorithms tailored to convert unreliable source text into professionally-sounding language. Human cognition is not optimized for this task.

Unfortunately, the stakes of LLM hallucinations are rather higher than discolored pixels and likely to increase as more individuals rely on LLMs with increasing autonomy, so-called agentic AI, to support their decision-making.$^{5}$ LLM's worldwide deployment and social media benefit from a business model predicated on the false premise that software is free when in fact users pay with their data subject to the promise to investors that high usage numbers will yield very high profits in future. Back in 2021, we highlighted concerns flagged to the US Senate in \textit{Testimony from a Facebook Whistleblower}$^{6}$. Since then, the CEO of Meta has apologized to the parents of children who had committed suicide, and a recent court ruling found Meta and YouTube guilty of building addictive social media platforms that harmed a 20-year old's mental health.$^{7}$ The problem is urgent and compounded by \textit{LLM slop}, which makes it easier than ever to populate social media platforms with unreliable, lowest common denominator or false content. 

\subsection{\textit{Humans are ``in the loop'', but are disempowered}}

Take, for example, the idea of using various LLMs in parallel to scale up the job of therapy advice, having a human therapist on hand who is alerted to \textit{take over} whenever an LLM produces an output from a pre-defined list of red-flagged outputs. The situation is similar to the case of a self-driving car where a human working remotely is expected to intervene to avoid a deadlock situation, or when a passenger in the driving seat has liability in the case of an accident with disregard for \textit{automation complacency} (as humans gain confidence in the system, their attention to the task at hand, e.g. driving, diminishes). As with the case of the therapist, here too the human is not empowered in the interaction with the system, but is made responsible when things go wrong. \textit{Human-in-the-loop} only works when humans are empowered by the interaction with the AI system and can intervene meaningfully in the decision-making process.

The seductive power of LLM capabilities introduces further risks of disempowerment. From students to politicians, LLM users are increasingly passing off AI-generated content as their own. European Union (EU) regulation requires developers to declare the role of AI in their tools. That does not apply to anyone using AI informally in the margins of organizational IT systems. How can we ensure, as an alternative to regulation, that people take responsibility for their decisions when they increasingly lean on AI to formulate them? Furthermore, how can we make individuals or organizations accountable in this new AI \textit{informal economy} where employees augment their work with LLMs in ways that are undisclosed or untracked? To be responsible for the work presupposes understanding the content! 

The highest risk (and arguably the highest opportunities) in AI are in agentic AI, that is, multiple LLMs with autonomy to take actions such as executing auto-generated code, collecting and processing data automatically to improve their own performance (known as recursive self-improvement), interacting or hacking into online systems, payment wallets, and web services to complete tasks - all based on the same unreliable foundation. Well-curated, mostly proprietary, high-quality data has become a very valuable asset recently to fine-tune or train domain-specific AI systems. Nevertheless, agentic AI continues to pose a serious cybersecurity risk to computer systems, with much of the latest research in this space having been focused on neuro-symbolic systems, LLMs with symbolic harnesses capable of providing assurances to various LLM outputs, e.g. using a symbolic theorem prover for the purpose of verification of auto-generated code.$^{8}$

\subsection{\textit{Systemic implications are far reaching}}
General purpose AI models have systemic implications that are less easy to predict and manage than the narrow, domain-specific applications that dominated the landscape in the late 2010s. Where models were largely developed within an industry or use case, such as healthcare, finance, or retail, the boundaries of those organizations and industries largely curtailed spillover consequences. 

Systemic implications today are far reaching, connecting into geopolitics and international development. The AI race, principally between the US and China, strains trade relations and introduces concerns about data sovereignty. The increased role of drones in modern warfare exacerbates pressures for autonomous AI capabilities. A dispute in 2026 between the company behind the \textit{Claude} LLM and the US Department of War risked a US\$200M contract, the dispute being on whether the company or the US government as its client should dictate the terms of use of Claude in relation to safeguards preventing use for mass surveillance. 

The economic disruption of AI is also being increasingly analyzed by governments and financial institutions, with particular concerns about LLMs replacing junior level staff, potentially disrupting the expertise transfer in the skills ladder, and about financial exposure due to circular financing and off balance sheet debt in AI companies.

Increasing levels of autonomy, such as AI systems with access to bank accounts, emails and calendars, introduce the possibility of machines operating as independent actors in society, with a range of risks to privacy and societal safety that would seem outlandish if it were not for the rapid progress of recent years. A small but growing number of AI users has been calling for independent rights and protections for AI systems, often grounded in a belief that AI systems are conscious or should be considered moral actors given their apparent agency. What will happen when these demands grow in volume and become the focus of political campaigns?

\subsection{Implications for the AI Accountability Ecosystem}

The AI accountability ecosystem argued that the then heavily imbalanced ecosystem failed to translate intensifying civil society concerns into productive changes and engender more positive attitudes towards AI. The then-ecosystem was skewed towards high-level principles (over 100 at the time) and disempowered tag-on external accountability mechanisms (such as senior boards of external observers). These external mechanisms proved unable to drive change internally, so the people involved raised high-profile public concerns instead, further alienating internal processes. 
An explicit ecosystem approach was suggested as a way to correct these issues, with attention given to developing multiple layers of interaction from actors and companies through to market counterparts, civil society, and governments. 

Many of the same issues apply even more now than five years ago. The traditional \textit{strong arm} of modern externally-led accountability seems to matter little in the face of LLM's technical, financial, and geopolitical reputation. In principle, society's strongest accountability systems are the independent members on corporate boards, government hearings, and political oversight. This does not seem to work in the case of frontier AI, as best described by a former OpenAI board member in a report calling for better mechanisms to manage the increasingly divergent incentives at play.$^{9}$

\section{Enhancing the AI Accountability Ecosystem}

Three interlinked adjustments can make the original ecosystem framework fit for purpose in the LLM era: supply chain orientation, outcomes monitoring, and user accountability. The revised framework is summarized in Figure 1.

\begin{figure}[h]
\centering
\includegraphics[width=1.2\textwidth, angle=90]{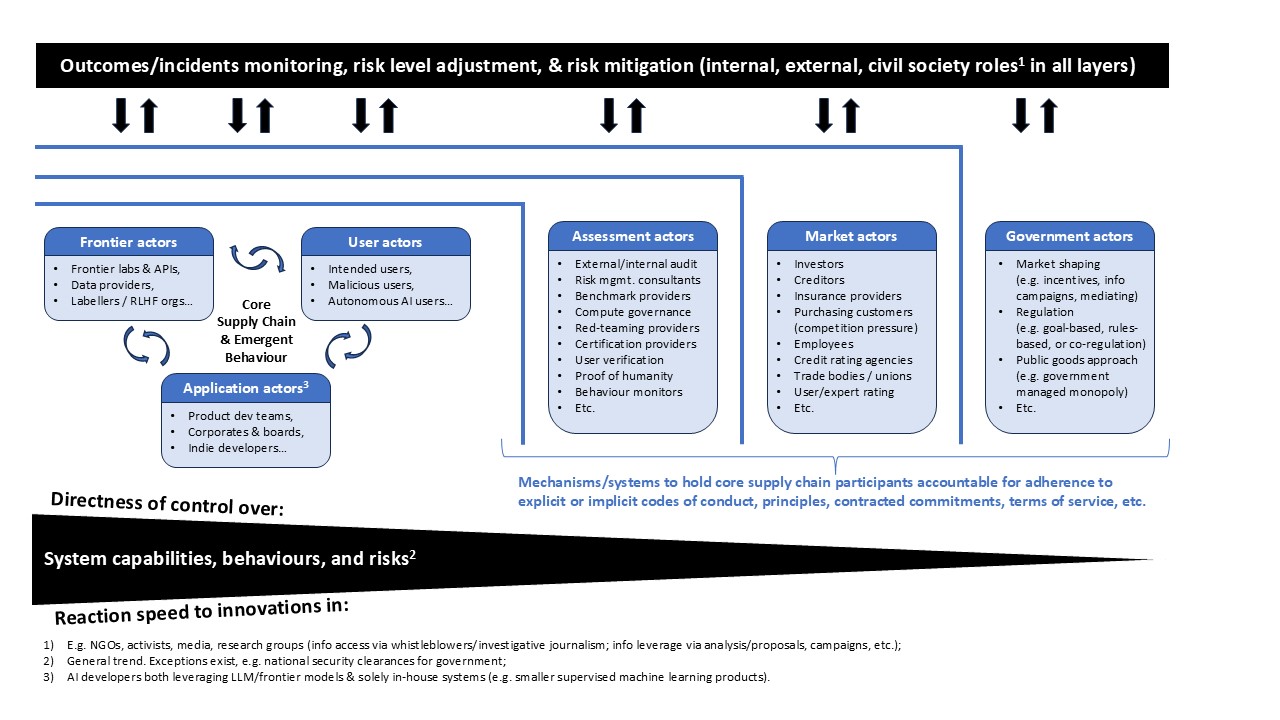}
\caption{The Accountability Ecosystem fit for LLMs (simplified schematic for a mixed market economy in a democracy). The framework identifies the core supply chain of AI systems as encompassing three sets of actors: frontier model developers, application developers, and users. Those involved in model assessment, market participants, and government actors form three consecutive layers that interact with the core supply chain, having access to multiple accountability mechanisms and systems to hold the core supply chain to account, but with increasingly reduced reaction speed and directness of control. Outcomes need to be monitored and risks managed for all six sets of actors in the AI ecosystem.}
\end{figure}

\subsection{\textit{Supply chain re-orientation}}


In the era of LLMs, centering accountability on a bounded \textit{actor} (a product or development team) as the unit of technical agency is no longer adequate. Instead, accountability should be reoriented to multi-layered AI supply chains.$^{10}$ The core supply chain has interactions among three sets of actors: frontier model developers, application developers, and users. While some AI applications are still built from scratch, an increasing proportion either require or incorporate LLM-related functionality from frontier labs. Application developers were the central \textit{actor} in the original proposal, but the core supply chain perspective translates responsibility into a distributed and interdependent phenomenon. 

Moving beyond the core supply chain, we identify assessment actors as including those dedicated to AI developers, such as benchmark designers and promoters and red-teaming providers, as well as pre-existing solutions that can be repurposed for AI, such as external and internal audit providers and risk management consultants. Market actors also continue to play an important role in holding core supply chain actors to account, ranging from investors, creditors, and insurance providers through to employees and trade bodies. This shift enables new forms of non-government accountability through assurance markets and standardized validation practices. One can require upstream actors to provide model documentation, eval results, and safety guarantees that downstream actors should rely on contractually. Independent auditors and benchmark providers can validate claims (e.g. robustness, bias) at different points in the stack. In this sense, accountability is composable: each layer is accountable for its contribution, and assurance propagates through the system via documentation, certification, and contractual obligations.

This approach aligns with emerging regulatory thinking (e.g. “value chain responsibility” in the EU AI Act), but is particularly powerful in non-regulatory contexts, where procurement requirements, platform policies, and third-party audits can enforce standards without direct state intervention. It also mirrors mature accountability systems in other domains (e.g. financial auditing, supply chain certification), suggesting a plausible pathway for scaling AI governance beyond firm-level self-regulation.

\subsection{\textit{Monitoring Outcomes}}

In the era of LLMs, accountability must be grounded in post-deployment, run-time monitoring of systems. This reflects the reality that many harms such as misuse and emergent behavior are not predictable at development time. As a result, accountability systems must incorporate mechanisms for (i) monitoring evolving requirements, (ii) detecting, reporting, and responding to failures often identified by users, civil society, or downstream integrators, and (iii) engaging with commissioned or external red-teaming investigations.

Recent literature also supports a shift towards continual accountability grounded in the monitoring of observed outcomes. A central finding in the International AI Safety Report 2026 is the existence of an \textit{evaluation gap}: systems often perform well in controlled pre-deployment testing but behave unpredictably in real-world contexts. This gap provides strong empirical reason to move beyond static audits to ongoing monitoring and feedback loops.$^{11}$

At the same time, policy and industry frameworks are converging on risk-tiered approaches. The International AI Safety Report and related governance initiatives emphasize that AI risks vary widely, from low-level inaccuracies to systemic or catastrophic harms, and that oversight mechanisms should scale accordingly. In parallel, technical and organizational frameworks 
increasingly emphasize risk classification, proportional controls, and adaptive mitigation strategies that should be mindful of the evolving capabilities and specific use cases of AI applications.$^{12}$

Outcomes and the associated risk-level assessment and risk management processes need to be monitored at all levels of the ecosystem. Frontier labs, developers, and users, as well as various assessment providers, market actors, and government systems, are all potential loci for risk origination or observation. 
The ecosystem perspective identifies a broad range of relevant actors. Frontier labs or the open source community might have the most rapid reaction speeds and direct control in most cases - and do bear important responsibility - but others cannot slough off all their duties to them. Outcomes monitoring across these layers would ideally leverage a mixture of formal systems that are mandated and funded to monitor outcomes, alongside informal, civil society and volunteer methods.

\subsection{\textit{User Accountability}}
In today's world, the end user or customer is also a focal point for accountability. Model users today are active participants in shaping AI system behavior, particularly in the context of LLM \textit{Chain of Thought} prompting. Recent research about AI risk highlights the growing importance of misuse by adversarial prompting and interactive vulnerabilities, including scams, manipulation, and cyber attacks facilitated by general-purpose models.$^{13}$

This creates a new class of accountability challenges that cannot be addressed solely through developer-side controls. Instead, accountability must be shared between system providers and users. In practice, this requires a combination of technical safeguards (e.g. formal verification, alignment techniques, refusal of certain requested behaviors), platform governance (e.g. usage policies, enforcement), and user-facing interventions (e.g. warnings, time-out, education and friction). In some contexts, contractual or legal mechanisms may also be used to assign responsibility for misuse. The AI industry cannot continue with terms of service requirements that are constantly being changed, rarely read, and even more rarely monitored, let alone enforced.

Accountability on professionals also gains new challenges as their AI tools grow in sophistication compared to previous generations of software. Clients will expect professionals, such as judges, accountants, or physicians, to  maintain accountability for their decisions. Such professionals will increasingly receive and review AI-generated recommendations. They will be expected to override inappropriate AI outputs, but will need to do so in a way that is documented, transparent, auditable. We can expect legal challenges from clients seeking to see the original AI advice, including cases where the AI advice turns out, in retrospect, to have been more appropriate than the revised professional decision. Training for professionals and new professional norms or guidelines will be important in navigating this transition, as well as technical progress on AI explainability, discussed below.

Importantly, incorporating user accountability does not imply shifting blame away from developers, but rather recognizing that AI systems are interactive and co-produced in use, especially agentic AI. Indeed, suitably autonomous agentic AI systems will also need to be held accountable, potentially using metrics that quantify the robustness of AI agent "identity".$^{14}$ This requires designing systems that are robust to misuse while also establishing clear norms and consequences for harmful user behavior. It also reinforces the importance of monitoring and feedback loops, since users are often the first to discover system vulnerabilities. 

\section{An illustration of the updated ecosystem}
We illustrate the importance of the three adjustments by comparing the original ecosystem with the adjusted ecosystem in a hypothetical scenario. 
Suppose a system is procured to support the identification of potential security risks through the analysis of publicly available communications. The system was not developed by a single actor but assembled across a delivery chain comprising a frontier AI model provider, a platform or API intermediary, a systems integrator, and a deploying agency. The resulting system is therefore best understood not as a bounded product but a socio-technical collaboration spanning multiple organizations and layers of control. Over time, several concerns emerge. First, the system is gradually repurposed beyond its original remit, expanding from targeted threat detection to broader population-level monitoring. Second, analysts discover that certain prompting strategies enable the model to generate inferences about individuals’ political affiliations or other sensitive attributes. 
Third, some users begin to develop informal workarounds that circumvent built-in safeguards, effectively \textit{jail-breaking} the system to produce outputs that would otherwise be restricted. 


Traditionally, accountability would be achieved through improved transparency, clearer documentation, and the progressive layering of internal and external audit mechanisms. However, here, the behaviors of concern emerge from interactions across the supply chain and system usage. As a result, internal audit mechanisms may lack visibility over upstream model behavior or downstream user practices, documentation may not capture the full range of emergent use cases, and external scrutiny may operate with limited access to the overall system. The likely outcome is a delayed and partial visibility of issues and a reactive pattern of accountability, characterized by incremental rather than systemic remediation.

By contrast, an accountability ecosystem reoriented around supply chains introduces a different set of mechanisms and, correspondingly, different dynamics of response. Responsibilities should be specified across the delivery stack. Example responsibilities are: including upstream model providers as accountable for documenting system capabilities, limitations, and known risk domains, platform providers as accountable for enforcing usage policies and monitoring access patterns, integrators for aligning system functionality to intended use cases, and deploying organizations for defining operational boundaries and oversight. 
These responsibilities are instantiated through contractual arrangements, benchmark validation, and assurance processes that enable claims to be tested and compared across actors. In this hypothetical case, an anomalous usage pattern indicative of population-level monitoring may be detected at the platform layer, known limitations in inferring sensitive attributes may have been documented by the model provider, and deviations from intended use may be identified by the integrator. Rather than a single actor bearing diffuse responsibility, accountability is distributed but structured, ideally to allow targeted interventions at the appropriate levels.

The incorporation of continual outcomes monitoring further differentiates the updated ecosystem from the original formulation. The updated approach places greater weight on the systematic observation of real-world system behavior. While it is challenging to engineer, we can imagine that signals from internal logs, user interactions, and external stakeholders might be aggregated and assessed through a monitoring function that supports a tiered risk management. 
In the original framework, users are primarily situated within the market or civil society layers, with limited emphasis on their role shaping system behavior. By contrast, the updated ecosystem recognizes that, in the context of LLMs, system outputs are co-produced through interaction. Adversarial prompting and the circumvention of safeguards are therefore not exceptional events but foreseeable behaviors that must be incorporated into accountability design. Mechanisms such as usage monitoring and access controls operate alongside technical safeguards to address the risks. Accountability thus shifts from a predominantly procedural exercise to a dynamic, evidence-based process.

Example success metrics would also be distributed across ecosystem layers. The core system should identify a proportion of risks that ultimately turn out to be validated as being genuine issues requiring interventions, given a particular tolerance. The number of risks flagged by the system, the number investigated, and subsequent outcomes can all be monitored. Automated transparency with anonymized metrics for the types of prompts and risks identified is expected to make it harder for users to re-purpose the system without the trend being noticed. An overall success metric is the number and severity of AI-related incidents that are reported, related to the system's scope and potential misuse, perhaps leveraging a formal version of the OECD's AIM database (AI Incidents and Hazards Monitor), which collates issues reported in diverse media.

Taken together, the three updates: distributed responsibility across the supply chain, continual outcomes monitoring, and explicit user accountability, lead to materially different patterns of accountability. 
None of this is straightforward to design, implement, or oversee, but having a vision for the accountability ecosystem is a prerequisite to success that does not depend on worldwide agreement on AI.

\section{On Explainability, Transparency, and Neurosymbolic AI}

The accountability ecosystem proposal from 2021 argued for the importance of explainable AI (XAI) as a tool for increasing reliability and trust in neural networks. How does this requirement fare five years on? Although XAI continues to be relevant to improving performance and trust in industry-specific AI, achieving explainability for entire LLMs, known as global XAI, is a very difficult, if not impossible task. This is because of the current scale of LLMs with billions of trained parameters (network weights). Efforts continue, however, trying to achieve local XAI and mechanistic interpretability (explaining individual cases, but struggling to scale to flexible questions across full models), and in the field of network \textit{distillation} to create simpler models, as was done by the Chinese DeepSeek LLM.$^{15}$

Research in the area of neurosymbolic AI continues to make progress as an alternative to the current LLM approach, advocating the use of something known as the neurosymbolic cycle of AI system development: train a small neural network with some available data, extract symbolic knowledge from that trained network using XAI, reasoning about what has been learned and evaluating results before further training with more data and parameters. This iterative development approach has been shown to empower domain experts in the interaction with AI systems during training, enabling user interventions to fix mistakes or improve performance, formal reasoning based on the extracted knowledge, and parsimony when knowledge is consolidated and reused across applications (contrast with the scaling of LLM training and its vast energy costs).$^{11}$ 

Using LLMs alongside formal proof assistants or knowledge graph queries are steps in the direction of neurosymbolic integration. The use of distillation in the deployment of the DeepSeek LLM, instigated by a lack of access to the latest graphics processing hardware in China, is also a step in the direction of a more parsimonious model as advocated by neurosymbolic AI, but only if the knowledge distillation is also leveraged for the purpose of explainability.


Standard approaches to software documentation are also coming under pressure, given the volume and complexity of AI-generated and AI-powered software. New approaches will be needed to maintain code transparency, to complement the model transparency that neurosymbolic techniques seek to bring. One promising idea is the continuous assurance framework \textit{Audit-as-Code} intending to map governance requirements into auditable rules.$^{16}$

\section{Conclusion and Future Work}

Multiple concerns motivate the need for an AI accountability ecosystem, including several highlighted in the latest AI index report 2026: ``AI models are achieving breakthrough results in science and complex reasoning, but at a concerning environmental toll. Today’s most capable modern models are now among the least transparent. Meanwhile, AI’s workforce disruption has moved from prediction to reality, hitting young workers first''.$^{17}$

This paper has argued that the accountability ecosystem for AI, as originally conceived in the context of machine learning prior to the release of LLMs for general public use, continues to be relevant but requires substantive adaptation to remain effective in the era of large language models. The general-purpose nature and widespread use of LLMs, their deployment through complex and multi-layered supply chains, and their susceptibility to emergent and interaction-driven behaviors challenge the assumption that accountability can be centered on a single organizational actor or addressed primarily through \textit{ex ante} processes.

We have argued that the accountability ecosystem approach continues to be relevant because AI cannot be seen as a \textit{ladder} from sub-human to human and super-human intelligence, but is rather an entire ecosystem of human and machine collaboration. As a result, accountability in AI needs to take this ecosystem into consideration, an idea that is aligned with the recently proposed \textit{Copernican view of AI}.$^{18}$ However, the original accountability ecosystem framework of 2021 requires updating to consider the new reality of the era of LLMs, with the amazing speed and investment scale of AI technology deployment. We therefore proposed three interrelated adjustments to the accountability ecosystem that should make it fit for purpose in the era of LLMs: first, accountability should be reoriented around the AI supply chain, recognizing that system behavior is co-produced across multiple layers of development, integration, and deployment. This shift enables a more appropriate allocation of responsibility and supports the development of assurance mechanisms such as benchmarking, auditing, certification, and contractual controls that operate across organizational boundaries; second, greater emphasis should be placed on outcomes monitoring and a tiered risk management, reflecting the limits of anticipatory governance in the face of uncertain or evolving system behaviors (continual observation of real-world performance coupled with structured escalation pathways should allow accountability to be grounded in evidence and adapted over time); third, the ecosystem should explicitly incorporate user accountability and safeguards, acknowledging that despite AI technology continuing to be centralized in Big Tech infrastructure, AI's impact is distributed across the ecosystem and, hence, AI accountability mechanisms should see users as active participants in shaping system inputs and outputs.

The hypothetical case presented in this paper seeks to illustrate how these adjustments can lead to materially different accountability dynamics. Under the earlier actor-centric model, the distributed nature of responsibility and the opacity of system behavior create conditions for delayed detection, contested attribution, and reactive intervention. By contrast, an ecosystem structured around supply chain accountability, continual outcomes monitoring, and user-level controls should enable earlier identification of issues, clearer allocation of responsibility, and better coordinated responses across system components. 

From a policy perspective, these findings highlight the increasing importance of non-regulatory and hybrid governance mechanisms. In particular, assurance markets comprising independent auditors, benchmark providers, certification bodies, and risk assessment services offer a pathway for translating high-level principles into operational practices without requiring immediate formal regulation. Similarly, procurement standards, contractual requirements, insurance mechanisms, and platform-level controls can exert significant influence over system design and deployment, particularly where they are embedded within supply chain relationships. These mechanisms are often more adaptive and responsive than formal regulation, and may be better suited to the pace and uncertainty associated with frontier AI development.

At the same time, the effectiveness of such non-regulatory interventions depends on a number of enabling conditions. These include the availability of credible and standardized evaluation methods, sufficient transparency and information-sharing across the supply chain, the alignment of incentives among developers, deployers of AI technology and users, and the presence of institutional actors capable of aggregating and acting on signals from monitoring systems. In the absence of these conditions, there is a risk that non-regulatory mechanisms become fragmented, performative, or captured by dominant actors. Moreover, certain categories of risk, particularly those relating to systemic harms, national security, or fundamental rights, may ultimately require formal regulatory intervention to ensure online safety, consistency and enforceability. 


Future work should explore how the accountability ecosystem can be operationalized, with four particular areas of focus. First, explaining what each actor within the AI accountability ecosystem is expected to do in practice and how responsibilities are allocated across the AI supply chain. Second, identifying suitable evaluation metrics or \textit{success criteria} that organizations can use to assess whether a governance framework has been implemented effectively or is achieving its intended objectives. Third, establishing how to manage ongoing changes in accountability mechanisms as AI systems continue to increase in capability and societal impact. Fourth, defining the concrete actions, workflows, or governance activities required of the various stakeholders involved in developing, deploying, using, and monitoring AI systems. 

As we hear of autonomous LLMs \textit{escaping the sandpit}, cyber attacks on AI platforms, and Big Tech leaders calling for slowing down AI development due to their risks, it is important to focus where targeted actions can make a difference. In a recent G7 meeting, Big Tech CEOs told world leaders that governments need to act on AI, and fast. However, beyond that, consensus collapsed. Some wanted a US-led coalition to control access and box out open-weight rivals. Others warned against concentrating too much power in the hands of a few companies. Others still proposed the creation of a new technical standards body responsible for carrying out AI model reviews prior to release. Three different governance philosophies, each aligned with their company's different revenue models, regulatory exposure and incentives. In this article, we do not ask governments to pick a geopolitical lane or referee a fight between AI labs. We call for something much simpler that does not require waiting for international consensus: accountability distributed across the supply chain with emphasis on outcomes monitoring and incorporating user accountability.

Computation abstractions have taken us from hand-welded hardware to assembly language, from early programming languages to object-orientation, dynamic typing in recent scripting languages, and application software that leverages massive standard libraries. Each step has enabled a phase transition in both the number of developers and the kinds of solutions they can generate. If we get the accountability ecosystem right, the LLM and agentic AI era might become just the next layer of computation abstraction, allowing users to focus where it really matters: the specification of systems' requirements and software validation protocols. With the right technical innovations, neurosymbolic AI and XAI tools, iterative automated workflows, and appropriate ecosystem accountability, the power of AI may become sustainably and safely available to a new generation of creatives, larger and faster than ever before.

\section*{ACKNOWLEDGMENTS}
The authors would like to thank Simo Dragicevic for initiating the thinking on AI accountability ecosystems within the gambling domain.

\def\refname{References}



\end{document}